\documentclass[10pt,aps,prl,twocolumn,showpacs,amssymb,floatfix]{revtex4}
\usepackage{amsmath,graphics,epsfig,mathrsfs}
\usepackage{bm}
\newcommand{\mac}{\mathcal}
\newcommand{\msc}{\mathscr}
\newcommand{\tx}{\text}
\newcommand{\ti}{\textit}

\newcommand{\nn}{\nonumber}
\newcommand{\pat}{\partial}
\newcommand{\pr}{\prime}
\newcommand{\para}{\parallel}

\newcommand{\raw}{\rightarrow}

\newcommand{\alp}{\alpha}
\newcommand{\dlt}{\delta}
\newcommand{\Dlt}{\Delta}

\newcommand{\eps}{\epsilon}

\newcommand{\og}{\omega}
\newcommand{\Og}{\Omega}
\newcommand{\sg}{\sigma}
\newcommand{\Sg}{\Sigma}

\newcommand{\etal}{{\em et al.,~}}
\newcommand{\ie}{{i.e.,~}}
\newcommand{\eg}{{e.g.,~}}

\begin{document}

\title{Diffusive Spin Dynamics in Ferromagnetic Thin Films with a Rashba Interaction}
\author{Xuhui Wang}
\email{xuhui.wang@kaust.edu.sa}
\author{Aurelien Manchon}
\email{aurelien.manchon@kaust.edu.sa}
\affiliation{King Abdullah University of Science and Technology (KAUST),
Physical Science and Engineering Division, Thuwal 23955-6900, Saudi Arabia}

\date{\today}
\begin{abstract}
In a ferromagnetic metal layer, the coupled charge and spin
diffusion equations are obtained in the presence of both Rashba
spin-orbit interaction and magnetism. The misalignment between the
magnetization and the nonequilibrium spin density induced by the
Rashba field gives rise to Rashba spin torque acting on the
ferromagnetic order parameter. In a general form, we find that the
Rashba torque consists of both in-plane and out-of-plane
components, \ie $\bm{T}=T_{\bot}\hat{\bm{y}}\times{\hat{\bm
m}}+T_{\para}{\hat{\bm m}}\times({\hat{\bm y}}\times{\hat{\bm
m}})$. Numerical simulations on a two-dimensional nanowire
consider the impact of diffusion on the Rashba torque and reveal a
large enhancement to the ratio $T_{\para}/T_{\bot}$ for thin
wires. Our theory provides an explanation for the mechanism driving
the magnetization switching in a single ferromagnet as observed in
the recent experiments.
\end{abstract}
\maketitle

The manipulation of spin degrees of freedom and spin-charge conversion are at the core of
the rapid developing field of spintronics \cite{zutic-rmp-2004}.
A semiconductor-based two-dimensional electron
gas (2DEG) lacking inversion symmetry is known to electrically generate
nonequilibrium spin density through the spin-orbit interaction \cite{dp-she-1971, edelstein-1989}.
The same type of spin-orbit interaction, named after Rashba \cite{bychkov-rashba-jetp-1984},
is the driving force behind numerous recent interesting discoveries,
\eg the well-known spin-Hall effect \cite{murakami-science-2004,sinova-prl-2004}.

\begin{figure}
\centering
\includegraphics[trim = 60mm 45mm 70mm 100mm, clip, scale=0.75]{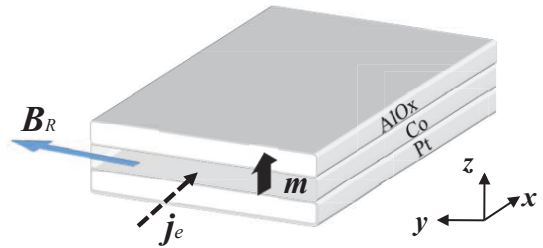}
\caption{\label{fig:setup} A schematic view of a Co film
sandwiched by a metal-oxide layer (AlO$_{x}$) and a heavy metal
(Pt). The magnetization direction $\hat{\bm{m}}$ is arbitrary, the
in-plane charge current is $\bm{j}_{e}$, and $\bm{B}_{R}$ is the
Rashba effective field perpendicular to $\bm{j}_{e}$.}
\end{figure}

Besides its dominating role in semiconductor structures,
Rashba spin-orbit interaction is expected to emerge in a thin ferromagnetic layer, such as cobalt (Co),
sandwiched asymmetrically between a heavy metal thin film (Pt) and a layer
of metal oxides (AlO$_{x}$) \cite{mihai1,pi,mihai3,mihai2}.
In such a quasi-two-dimensional metal layer (see Fig.\ref{fig:setup}),
the effective field $\bm{B}_{R}$ generated by
the spin-orbit interaction is predicted to excite the ferromagnetic order
parameter by a charge current \cite{manchon-prb,others,tatara},
which has been confirmed by several experiments \cite{mihai1,pi,mihai3,mihai2}.
This spin torque, coined Rashba torque, falls into to a
broader family where the spin-orbit interaction enables
a transfer of angular momentum between the spin and orbital degrees of freedom,
and has been observed in diluted magnetic semiconductors \cite{chernyshov,suzuki,fang}.
The same type of spin-orbit-induced torque is predicted to improve current-driven domain wall
motion \cite{tatara,others} as it has been shown experimentally \cite{mihai2}.
Recently, Miron \etal \cite{mihai3} has demonstrated the current-induced magnetization switching in
a \ti{single} ferromagnet, which represents an outstanding alternative to the celebrated
Slonczewski-Berger spin-transfer torque \cite{slonczewki-jmmm-1996,berger-prb-1996}
that requires noncollinear magnetic textures such as spin valves or domain walls \cite{refstt}.

In a nonmagnetic 2DEG, the diffusive spin dynamics in the presence of
Rashba spin-orbit interaction and D'yakonov-Perel spin
relaxation \cite{dp} has attracted significant attention
\cite{mishchenko-halperin-prb-2003,burkov-prb-2004,mishchenko-prl-2004},
resulting in further intrigue such as spin-Hall edges \cite{adagideli-bauer-prl-2005}.
We foresee that in a metallic ferromagnetic layer accommodating both
a Rashba spin-orbit interaction and an exchange splitting, the competition
between spin relaxation (induced by random magnetic impurities and D'yakonov-Perel)
and the spin precession enforced by both exchange and Rashba field
gives rise to complex spin dynamics that is important to
current-driven magnetization manipulation and anomalous Hall effect.
Meanwhile, theoretical accounts on Rashba torque are, so far,
limited to an infinite medium \cite{manchon-prb,others}, where diffusive
motion is absent. In this Letter, we employ the Keldysh technique \cite{rammer-smith-rmp-1986}
to derive a diffusion equation describing the coupled dynamics of charge
and spin in a quasi-two-dimensional ferromagnetic layer submitted
to a Rashba spin-orbit interaction. We show that the coupling
between the magnetism and the spin-orbit interaction, as mediated by the
electrons through $s-d$ exchange, leads to a Rashba torque
($\bm{T}$) that has both out-of-plane and in-plane components, \ie
$\bm{T}=T_{\bot}\hat{\bm{y}}\times{\hat{\bm m}}
+T_{\para}{\hat{\bm m}}\times({\hat{\bm y}}\times{\hat{\bm m}})$.
In the case of a narrow magnetic wire, we show that the in-plane torque $T_{\para}$ can
be enhanced significantly.

The total Hamiltonian for conducting electrons
(with effective mass $m$) in a quasi-two-dimensional ferromagnetic layer
in $x-y$ plane is ($\hbar=1$ is assumed throughout)
\begin{equation}
H_{T}=\frac{{\hat{\bm k}}^2}{2m}
+\alp{\hat{\bm\sg}}\cdot({\hat{\bm k}}\times{\hat{\bm z}})
+\frac{1}{2}\Dlt_{xc}{\hat{\bm\sg}}\cdot{\hat{\bm m}}+H^{i}
\label{eq:total-hami}
\end{equation}
where $\bm{k}$ is the momentum, $\alp$ is the Rashba constant, $\hat{\bm\sg}$ the
Pauli matrices, $\Dlt_{xc}$ the ferromagnetic exchange splitting,
$\hat{\bm{m}}$ the magnetization direction,
and $H^{i}=\sum_{j=1}^{N}V(\bm{r}-\bm{R}_{j})$ the
spin-independent impurity potential with centers located at $\bm{R}_{j}$.
An electric field is applied and to be included in the following discourse.
Using Dyson equation \cite{rammer-smith-rmp-1986},
the advanced and retarded Green's functions $\hat{G}^{A}$ and $\hat{G}^{R}$,
Keldysh function $\hat{G}^{K}$, and the self-energy $\hat{\Sg}^{A,R,K}$
are related by the quantum kinetic equation
\cite{rammer-smith-rmp-1986, mishchenko-prl-2004}
\begin{align}
[\hat{G}^{R}]^{-1}\hat{G}^{K}-\hat{G}^{K}[\hat{G}^{A}]^{-1}
=\hat{\Sg}^{K}\hat{G}^{A}-\hat{G}^{R}\hat{\Sg}^{K},
\label{eq:kinetic-equation}
\end{align}
where all Green's functions are the full functions.

To obtain a diffusion equation from Eq.(\ref{eq:kinetic-equation}), we assume
short-range $\dlt$-function type impurity scatterers at low concentration with a weak
coupling to electrons \cite{mishchenko-prl-2004}, then a second-order Born approximation
is justified, \ie
$\hat{\Sg}^{A,R,K}(\bm{r},\bm{r}^{\pr})
=\dlt(\bm{r},\bm{r}^{\pr})\hat{G}^{A,R,K}(\bm{r},\bm{r})/(m\tau)$.
The momentum relaxation rate $1/\tau$
due to spin-independent impurities is evaluated at the Fermi energy.
The quasiclassical distribution function
$\hat{g}\equiv\hat{g}_{\bm{k},\eps}(T,\bm{R})$, as the Wigner transform of the Keldysh function
$\hat{G}^{K}(\bm{r},t;\bm{r}^{\pr},t^{\pr})$, is obtained
by integrating out the relative spatial-temporal coordinates while retaining the
center-of-mass ones $\bm{R}=(\bm{r}+\bm{r}^{\pr})/2$ and
$T=(t+t^{\pr})/2$. A gradient expansion on both sides of
Eq.(\ref{eq:kinetic-equation}) followed by a Fourier transform in time domain
gives us
\begin{align}
\Og\hat{g}-b_{k}[\hat{U},\hat{g}]
&= -\frac{i}{2}\left\{\frac{\bm{k}}{m}
+\alp(\hat{\bm{z}}\times\hat{\bm{\sg}}),\nabla_{\bm{r}}\hat{g}\right\}\nn\\
&-\frac{1}{\tau}\left[\hat{G}^{R}(\bm{k},\eps)\hat{\rho}(\eps)-\hat{\rho}(\eps)\hat{G}^{A}(\bm{k},\eps)\right],
\label{eq:qke-2}
\end{align}
where $\hat{\rho}$ is the density matrix. In Eq.(\ref{eq:qke-2}),
$\Og\equiv\og+i/\tau$, $b_{k}\equiv|\Dlt_{xc}\bm{m}/2+\alp\bm{k}\times\hat{\bm{z}}|$,
the operator $\hat{U}\equiv\hat{\bm{\sg}}\cdot(\Dlt_{xc}\bm{m}/2+\alp\bm{k}\times\hat{\bm{z}})/b_{k}$,
and $\{\cdot,\cdot\}$ denotes the anticommutator.

We solve Eq.(\ref{eq:qke-2}) according to the discussion outlined in Ref.\cite{mishchenko-prl-2004}:
find a formal solution $\hat{g}=\msc{L}[\hat{K}^{(0)}+\hat{K}^{(1)}]$ in terms of
the lowest order approximation
\begin{align}
\hat{K}^{(0)}=& \frac{i}{\tau}\left[\hat{G}^{R}(\bm{k},\eps)\hat{\rho}(\eps)
-\hat{\rho}(\eps)\hat{G}^{A}(\bm{k},\eps)\right]
\end{align}
and a higher-order gradient correction
\begin{align}
\hat{K}^{(1)}=& -\frac{1}{2}\left\{\frac{\bm{k}}{m}
+\alp(\hat{\bm{z}}\times\hat{\bm{\sg}}),\nabla_{\bm{r}}\hat{g}\right\}
\end{align}
that is accounted by perturbation.
We then apply the formal solution in $\hat{K}^{(0)}$ to obtain the zeroth
order approximation that is to be substituted into $\hat{K}^{(1)}$ to obtain a
gradient correction, thus arriving at the first order approximation to $\hat{g}$.
We repeat this procedure to the second order for $\hat{g}$ \cite{mishchenko-prl-2004}.
This gradient expansion scheme is applicable as long as the
spatial gradient of the quasiclassical distribution
function is smooth at the scale of Fermi wavelength, i.e. $\pat_{r}\ll k_{F}$.

After angle averaging in momentum space and Fourier transforming
the second-order approximation back to real time,
we have a diffusion-type equation for the density matrix
\begin{align}
&\frac{\pat}{\pat t}\hat{\rho}+\frac{1}{\tau_{xc}}{\hat \rho}
-\frac{1}{2\tau_{xc}}({\hat{\bm z}}\times{\hat{\bm\sigma}})\cdot{\hat \rho}({\hat{\bm z}}\times{\hat{\bm\sigma}})
+\frac{1}{2 T_{xc}}(\hat{\sg}_{m}{\hat \rho}\hat{\sg}_{m}-{\hat \rho})\nn\\
&=D\nabla^{2}\hat{\rho}
+i C[{\hat{\bm z}}\times{\hat{\bm\sigma}},{\bm\nabla}{\hat \rho}]
- B \{{\hat{\bm z}}\times{\hat{\bm\sigma}},{\bm\nabla}{\hat \rho}\}
-i{\tilde\Delta}_{xc}[\hat{\sg}_{m},{\hat \rho}]\nn\\
&+\Gamma [({\hat{\bm m}}\times{\bm\nabla})_z{\hat\rho}
-\hat{\sg}_{m}{\bm\nabla}{\hat\rho}\cdot({\hat{\bm z}}\times{\hat{\bm\sigma}})
-({\hat{\bm z}}\times{\hat{\bm\sigma}})\cdot{\bm \nabla}{\hat\rho}\hat{\sg}_{m}]\nn\\
&-2 R \{\hat{\sg}_{m},({\hat{\bm m}}\times{\bm\nabla})_z{\hat \rho}\}
\label{eq:density-matrix-with-exchange}
\end{align}
where $\hat{\sg}_{m}\equiv \hat{\bm\sigma}\cdot\hat {\bm m}$
and we concentrate on quantities at Fermi energy.
The diffusion constant is $D=\tau v_{F}^{2}/2$,
given $v_{F}$ the Fermi velocity.
$\tilde{\Delta}_{xc}=(\Delta_{xc}/2)/(4\xi^2+1)$ where
$\xi^{2}=(\Dlt_{xc}^{2}/4+\alp^{2}k_{F}^{2})\tau^{2}$.
The other parameters in Eq.(\ref{eq:density-matrix-with-exchange}) are
\begin{align}
& C=\frac{\alp k_{F}v_{F}\tau}{(4\xi^{2}+1)^{2}},\;
\Gamma=\frac{\alpha\Delta_{xc}v_Fk_F\tau^2}{2(4\xi^2+1)^2},\;
R =\frac{\alp\Dlt_{xc}^{2}\tau^{2}}{2(4\xi^{2}+1)}\nn\\
& ~\frac{1}{\tau_{xc}}=\frac{2\alp^{2}k_{F}^{2}\tau}{4\xi^{2}+1},
~\frac{1}{T_{xc}}= \frac{\Dlt_{xc}^{2}\tau}{4\xi^{2}+1},\;
B =\frac{2\alpha^3k_F^2\tau^2}{4\xi^2+1},\nn
\end{align}
where we identify $\tau_{xc}$ as the D'yakonov-Perel relaxation
time. Equation (\ref{eq:density-matrix-with-exchange}) is valid in the dirty
limit (i.e. $\xi\ll 1$), which permits the approximation
$1+4\xi^{2}\approx 1$ throughout the following discussion.
We decompose the density matrix as
$\hat{\rho}=n/2+\bm{S}\cdot{\hat{\bm\sg}}$ to introduce the charge density $n$
and the spin density $\bm{S}$.
Spin transport in ferromagnetic layers in a real experimental setup \cite{mihai1,mihai2,mihai3}
is exposed to random magnetic scatterers, for which
an isotropic spin-flip relaxation $\bm{S}/\tau_{sf}$ is introduced phenomenologically.
In total, we have
\begin{align}
\frac{\pat n}{\pat t} =& D\nabla^{2}n+ B {\bm\nabla}_z\cdot{\bm S}\nn\\
& +\Gamma {\bm\nabla}_z\cdot{\hat{\bm m}}n
+R {\bm\nabla}_z\cdot{\hat{\bm m}}({\bm S}\cdot{\hat{\bm m}}),
\label{eq:charge-diffusion}\\
\frac{\pat \bm {S}}{\pat t} =& D \nabla^{2}{\bm S}
-\frac{1}{\tau_{\para}}\bm{S}_{\para}
-\frac{1}{\tau_{\perp}}\bm{S}_{\perp}\nn\\
&-\Dlt_{xc}\bm{S}\times\hat{\bm{m}}
-\frac{1}{T_{xc}}\hat{\bm{m}}\times(\bm{S}\times\hat{\bm{m}})\nn\\
&+B {\bm\nabla}_z n
+2 C {\bm\nabla}_z\times{\bm S}
+2 R (\hat{\bm{m}}\cdot{\bm\nabla}_z n)\hat{\bm{ m}}\nn\\
&+\Gamma \left[{\hat{\bm m}}\times({\bm\nabla}_z\times{\bm S})
+{\bm\nabla}_z\times({\hat{\bm m}}\times{\bm S})\right],
\label{eq:spin-diffusion}
\end{align}
where $\bm{\nabla}_z\equiv\hat{\bm{z}}\times\bm{\nabla}$.
Accounting for both D'yakonov-Perel and random magnetic impurities,
rate $1/\tau_{\para}\equiv 1/\tau_{xc}+1/\tau_{sf}$
measures the relaxation of the spin density
${\bm S}_{\para}\equiv S_x\hat{\bm{x}}+S_y\hat{\bm {y}}$,
as $1/\tau_{\perp}\equiv 2/\tau_{xc}+1/\tau_{sf}$
does to $\bm {S}_{\perp}\equiv S_z \hat{\bm {z}}$.
Parameter $T_{xc}$ sets a time scale for the
decay of the transverse (to $\hat{\bm{m}}$) component of
the spin density, thus contributing directly to the
spin torque \cite{stt-theory}.

For a broad range of the ratio $\alp k_{F}/\Delta_{xc}$,
Eqs.(\ref{eq:charge-diffusion}) and (\ref{eq:spin-diffusion})
govern the full spin dynamics in a ferromagnetic layer and constitute the main result
of this Letter. We can readily show \cite{wang-unpub-2011} that the
absence of magnetism ($\Dlt_{xc}=0$) leaves the $B$ term a source generating
spin density electrically \cite{edelstein-1989,mishchenko-prl-2004}.
When spin-orbit coupling vanishes ($\alpha=0$),
the first two lines in Eq.(\ref{eq:spin-diffusion}) survive to
describe a diffusive motion of spin density in a ferromagnetic metal,
which agrees excellently with early results
in the corresponding limit \cite{wang-unpub-2011, zhangli}.
The $C$ term represents a coherent precession of the
spin density around the effective Rashba field.
The precession of the spin density (induced by the Rashba field) around the exchange
field is described by the $\Gamma$ term, is thus at a higher order (compared to
$C$) in the dirty limit for $\Gamma=\Dlt_{xc}\tau C/2$. The $R$-term contributes to a
magnetization renormalization.

To illustrate the spin dynamics and the Rashba spin torque
embedded in Eq.(\ref{eq:spin-diffusion}),
we restore the electric field of strength $\msc{E}$ applied
along the $\hat{\bm{x}}$ direction
by a shift $\bm{\nabla}\raw\bm {\nabla}+e \msc{E} \hat{\bm{x}}\pat_\eps$
\cite{mishchenko-prl-2004}.
This Letter is focusing on a weak spin-orbit coupling
$\alpha k_F< \Dlt_{xc}$ \cite{manchon-prb,others};
\ie the spin density is aligned dominantly along the local magnetization.
Thus, the deviation due to effective Rashba field $\bm{B}_{R}\propto \bm{j}_{e}\times\hat{\bm{z}}$
(along the $\hat{\bm{y}}$ direction in the present setup) is considered as a perturbation.
We may approximate the energy derivative
$\pat_{\eps}\bm{S} \approx P_{F}\mac{N}_{F}\hat{\bm{m}}$
and $\pat_{\eps} n \approx n_{F}/\eps_{F}=\mac{N}_{F}$, given $\mac{N}_{F}$ the density
of states and $P_{F}$ the polarization; both quantities are at Fermi energy $\eps_{F}$.

Consider an infinite homogeneous ferromagnetic layer
\cite{manchon-prb}, we replace $\bm{\nabla}$ by
$e\msc{E}\hat{\bm{x}}\pat_{\eps}$ and Eq. (\ref{eq:spin-diffusion}) is
\begin{align}
\frac{\pat {\bm S}}{\pat t}=&-\frac{1}{\tau_{sf}}{\bm S}
-\frac{1}{T_{xc}}\hat{\bm{m}}\times({\bm S}\times\hat{\bm{m}})
-\frac{1}{\tau_{\Dlt}}\bm{S}\times\hat{\bm m}\nn\\
&+e \msc{E} \mac{N}_F\left[2 P_{F} C {\hat{\bm y}}\times{\hat{\bm m}}
+P_{F}\Gamma {\hat{\bm m}}\times({\hat{\bm y}}\times{\hat {\bm m}})\right].\label{eq:spin-diffusion2}
\end{align}
where the higher-order contribution proportional to $R$ is discarded
and $\tau_{\Dlt}\equiv 1/\Dlt_{xc}$ characterizes the time scale
of the precession of the spin density around the magnetization.
The spin torque exerted on the local magnetization
by the nonequilibrium spin density is given by
\begin{align}
\bm{T}=\frac{1}{\tau_{\Dlt}}\bm{S}\times\hat{\bm{m}}
+\frac{1}{T_{xc}}\hat{\bm{m}}\times({\bm S}\times\hat{\bm{m}}),
\end{align}
taking into account a fieldlike spin precession
and dephasing of the transverse component that is essentially the
Slonczewski-Berger type spin-transfer torque.
At a stationary state $\pat\bm{S}/\pat t=0$, Eq.(\ref{eq:spin-diffusion2})
leads to a general form
$\bm{T}=T_{\perp}\hat{\bm{ y}}\times\hat{\bm{m}}
+T_{\para}\hat{\bm{m}}\times(\hat{\bm{y}}\times\hat{\bm{m}})$ that is
\begin{align}
\bm{T}=\frac{j_e}{eD}\frac{P_{F}}{1+\zeta^2}
&\left[(2\eta C +\beta\Gamma){\hat{\bm y}}\times{\hat{\bm m}}\right.\nn\\
&\left.+(\eta\Gamma-2\beta C)\hat{\bm{m}}\times({\hat{\bm y}}\times{\hat{\bm m}})\right],
\label{eq:torque}
\end{align}
where $j_e=e^2 n_F\tau \msc{E}/m$ is the current density,
$\zeta=\tau_{\Dlt}(1/\tau_{sf}+1/T_{xc})$, $\eta=1+\zeta\tau_{\Dlt}/T_{xc}$, and
$\beta=\tau_{\Dlt}/\tau_{sf}$ \cite{zhangli}.
The first term in Eq.(\ref{eq:torque}) is an \ti{out-of-plane} torque
driven by the effective field $\bm{B}_{R}$ and is perpendicular to the
($\hat{\bm{m}}$, $\bm{B}_{R}$) plane.
Using the data from experiments \cite{mihai1}, we take
$\alp\approx 10^{-10}~\tx{eV}\tx{m}$, polarization $P_{F}=0.5$,
saturation magnetization $M_{s}\approx 1.09\times 10^{6}~\tx{A}~\tx{m}^{-1}$,
and the current density $j_{e}=10^{8}~\tx{A}~\tx{cm}^{-2}$, the effective field
generating the \ti{out-of-plane} torque
is estimated to be of the order of 1 T.

The second \ti{in-plane} torque arises from
the change of the spin density induced by its precession
around the exchange field ($\propto\Gamma$) as well as to the presence of
spin flip in the layer ($\propto\beta$), inducing a spin density
that is parallel to $\bm{B}_{R}\times\hat{\bm{m}}$ (or $\hat{\bm{y}}\times\hat{\bm{m}}$)
\cite{wang-unpub-2011}, hence leading to a
torque in the ($\hat{\bm{m}},\hat{\bm{y}}$) plane. In other words,
the in-plane torque is driven by an effective field along the
direction $\bm{B}_{R}\times\hat{\bm{m}}$.
In the limit $\tau_{sf}\raw \infty$,
the ratio $T_{\para}/T_{\perp}\approx \Dlt_{xc}\tau/4$, approaches
a constant.
When considering $T_{xc}\raw\infty$ and $\Gamma\ll C $ (dirty limit),
the in-plane torque diminishes and Eq.(\ref{eq:torque}) is in a
good agreement with Ref. [\onlinecite{manchon-prb}].

The impact of diffusion on the
Rashba spin torque is appreciated by numerically solving Eq. (\ref{eq:spin-diffusion})
in a semi-infinite ferromagnetic wire with a width $L$ along the $y$ axis.
The wire is open on the transport direction
$\hat{\bm{x}}$, along which an electric field is applied
(thus $\nabla_{x}\raw e\msc{E}\hat{\bm{x}}\pat_\eps$).
To compare with the recent experiment \cite{mihai3},
the magnetization is aligned perpendicularly along ${\hat{\bm z}}$.
Two boundary conditions are available in the transversal direction ($y$):
vanishing spin density ${\bm S}(y=\pm L/2)=0$ or
vanishing spin current $\mac{J}_y(y=\pm L/2)=0$ at the boundaries.
The former condition indicates an absorbent surface where spin-flip relaxation at the
boundaries is fairly strong, thus suppressing the spin density.
A vanishing spin current designates a fully reflective boundary
where the spin current is reflected. In the present simulation, we choose to adopt
the boundary condition according to the work of Kato \etal \cite{kato}:
optical detection on the spatial profile of the spin density induced by
the spin-Hall effect in a $n$-GaAs 2DEG reveals that the spin density
vanishes at the interfaces. We believe that the recent experiments on Rashba
spin torque in ferromagnetic layers \cite{mihai1,mihai3,mihai2} fall into this picture.

\begin{figure}
\centering
\includegraphics[trim = 0mm 0mm 0mm 0mm, clip, scale=0.38]{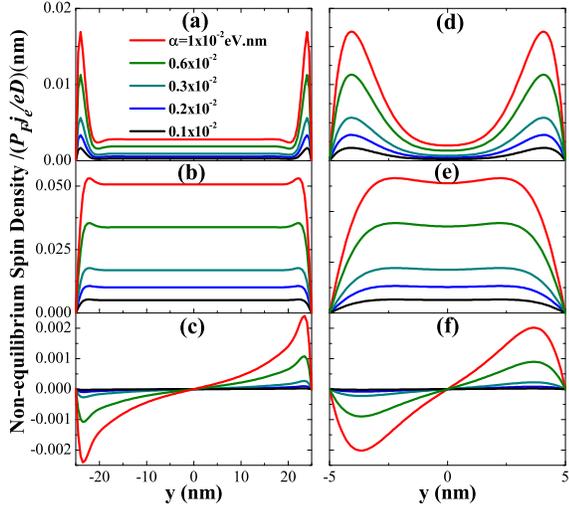}
\caption{\label{fig:Fig1}(Color online) Spatial profile of nonequilibrium spin density
$S_x$ (a),(d), $S_y$ (b),(e), and $S_z$ (c),(f) for a width $L=50$nm (a),(b),(c)
and $L=10$nm (d),(e),(f) for $\alpha=0.1-1\times10^{-2}$ eV$\cdot$nm.
The parameters are $\tau=10^{-15}$s, exchange splitting $\Dlt_{xc}=10^{14}$s$^{-1}$,
spin relaxation time $\tau_{sf}=10^{-12}$s,
diffusion constant $D=125\times10^{-6}$m$^{2}$~s$^{-1}$.}
\end{figure}

Figure \ref{fig:Fig1} shows the spatial profile of the three components
of the spin density: $S_x$ (panels a and d), $S_y$ (panels b and e), and $S_z$
(panels c and f), for various magnitudes of $\alpha$.
The spin-orbit interaction drives the spin dynamics in two prominent ways:
the first is explained by Edelstein \cite{edelstein-1989}; \ie
a nonequilibrium spin density at $\hat{\bm{y}}$ direction
is generated electrically, becoming a source to a
nonvanishing $S_{y}$ in the center of the wire.
Second, the spin density precesses around the total field
combining the exchange ($\Dlt_{xc}{\hat {\bm m}}\times{\bm S}$)
and the effective fields ($2 C {\bm\nabla}_z\times{\bm S}$).
In the case of a wide magnetic wire, this precession
is localized near the edges of the wire and vanishes
towards the center where the nonequilibrium
spin density generated by spin-orbit coupling
gets robust along the Rashba field ($\hat{\bm{y}}$ direction).
Consequently, the $S_{x}$ component,
though much smaller in magnitude than $S_{y}$, is peaked at
the boundaries; see Figs.\ref{fig:Fig1} (b) and \ref{fig:Fig1}(c).
The profile of the $S_{z}$ component has an opposite sign at two
edges, which is driven by the rotation of the spin accumulation
around the effective Rashba fields (pointing oppositely to each other)
generated while diffusing transversally along opposite directions
(\ie negative $y$ and positive $y$ axes).

In the case of a weak spin-orbit coupling addressed in
this Letter, the spin density profile is mainly
symmetric along the (transverse) $y$ direction.
In the opposite limit of a strong spin-orbit coupling, the spin-Hall effect
induces a large spin imbalance along the transverse direction of the wire,
which favors a deeper asymmetry in the spatial profile (not shown here).
The narrow width (shorter than the spin-flip diffusion length defined by $\tau_{sf}$)
of the wire prevents the relaxation of the spin density such
that its effective magnitude in the middle of the wire may differ from
its value in an infinite medium.

\begin{figure}
\centering
\includegraphics[trim = 0mm 0mm 0mm 0mm, clip, scale=0.42]{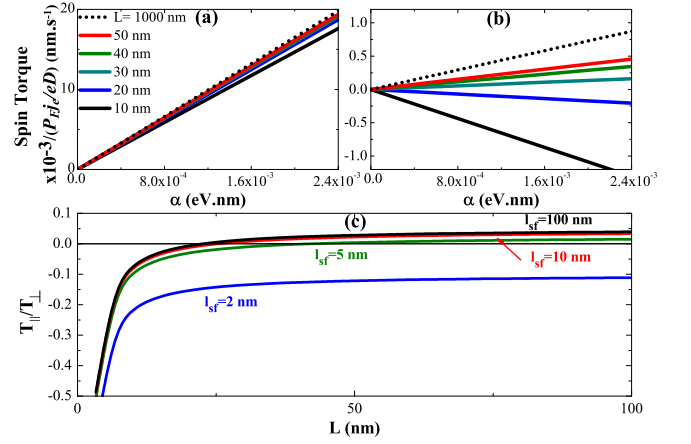}
\caption{\label{fig:Fig2}
(Color online) Out-of-plane (a) and in-plane (b) torques as a function of $\alp$
for different wire widths.
(c) Ratio $T_{\para}/T_{\bot}$ as a function of the wire width for different
spin diffusion lengths. The other parameters are the same as in Fig. \ref{fig:Fig1}.}
\end{figure}

Therefore, one expects a strong influence of the wire width
on the magnitude and sign of the
Rashba spin torque. Figures \ref{fig:Fig2}(a) and \ref{fig:Fig2}(b) display the magnitude of the
average torque $\int_{-L/2}^{L/2}\bm{T}dy/L$ at different wire widths as a
function of $\alp$. For a fixed $\alpha$,
by decreasing the wire width the in-plane torque $T_{\para}$ changes
its sign and the magnitude can be enhanced dramatically, while
the magnitude of the out-of-plane torque $T_{\bot}$ is just
moderately reduced.
Figure \ref{fig:Fig2}(c) shows the ratio $T_{\para}/T_{\bot}$ as a function of
the wire width for different spin diffusion lengths.
Several features deserve attention.
(i) For a given spin-flip relaxation time (therefore the spin diffusion length),
the ratio approaches a constant when increasing the width of the wire,
while its absolute value increases exponentially when decreasing the width.
(ii) For a wide wire with a large spin diffusion length,
the ratio tends towards the bulk value $T_{\para}/T_{\perp}=\Dlt_{xc}\tau/4$.
Using the present value $\Dlt_{xc}=10^{14}~\tx{s}^{-1}$ and $\tau=10^{-15}~\tx{s}$,
the bulk value is about $0.03$, which agrees well with the numeric
value (black solid line) in Fig.\ref{fig:Fig2} (c).
(iii) The sign of the ratio can be switched by the wire
width or the spin relaxation rate, as is reflected in
the second term in Eq. (\ref{eq:torque}).

In a recent work, Miron \etal \cite{mihai3} has demonstrated
a current-induced magnetization reversal
in a perpendicularly magnetized single ferromagnet.
In Eq.(\ref{eq:torque}) of this Letter, the effective field
(producing the in-plane torque)
parallel to $\bm{B}_{R}\times\hat{\bm{m}}$ explains
the effective perpendicular magnetic field $\bm{B}_{S_{z}}$
required to accomplish the magnetization switching \cite{mihai3}.
In the Co layer as sandwiched by metal oxides and heavy metal with a large spin-orbit
coupling, the spin-flip relaxation length is believed to be
much shorter than its value in the bulk Co, hence further enhancing
the in-plane torque: see Fig.\ref{fig:Fig2} (c).
If we take $\Dlt_{xc}\approx 5\times 10^{14}~\tx{s}^{-1}$
(or $0.3~\tx{eV}$) \cite{exchange-thin-film}, $\tau_{sf}\approx 10^{-12}~\tx{s}$,
and $\tau\approx 10^{-15}~\tx{s}$, then we have $\beta\approx 0.002$,
$\zeta\approx \tau_{\Dlt}/T_{xc}\approx 0.5$, and $\eta$ is of order 1.
A quick estimate suggests that the perpendicular switching field
generating the \ti{in-plane} torque is about 100 mT,
which agrees with the estimation in Ref.[\onlinecite{mihai3}].
In the case of magnetic domain walls,
the present torque acts like a transverse field
that can increase the Walker breakdown limit and
enhance the range for the nonadiabatic current-driven
domain wall motion \cite{tdw}. Miron \etal exploited this
characteristic to interpret large current-driven
domain wall velocities in perpendicularly magnetized domain walls \cite{mihai2}.

Most of the previous works on transitions metals have been carried out in
fairly complex structures where interfacial Rashba spin-orbit coupling has not been
evaluated experimentally or theoretically.
Identifying the relevant mechanisms is an urgent need \cite{cornell}.
Among them, the microscopic description of interfacial Rashba
spin-orbit coupling in ultrathin layers as well as the spin
dynamics in the presence spin-orbit coupling will provide essential information to the
electrical manipulation of spins in low dimensional systems.

We are grateful to G. E. W. Bauer, K. -J. Lee, J. Sinova, M. D. Stiles, X. Waintal,
and S. Zhang for stimulating discussions.


\begin{thebibliography} {999}
\bibitem{exchange-thin-film}
The exchange splitting in the bulk of Co is typically of order 1 eV and we expect
a lower value in the ultrathin film.
\end{thebibliography}
\end{document}